# Freely-suspended matrix-free metamaterials showing THz left-handed pass bands

H.O. Moser[1], J.A. Kong[2,3], L.K. Jian[1], H.S. Chen[2,3], G. Liu[1], M. Bahou[1], S.M.P. Kalaiselvi[1], S.M. Maniam[1], X.X. Cheng[3], B.I. Wu[2], P.D. Gu[1], A. Chen[1], S.P. Heussler[1], Shahrain bin Mahmood[1], L. Wen[1]

[1]*Singapore Synchrotron Light Source (SSLS), National University of Singapore (NUS), 5 Research Link, Singapore 117603*

[2]*Research Laboratory of Electronics, Massachusetts Institute of Technology, Cambridge, Massachusetts 02139, USA*

[3]*The Electromagnetics Academy at Zhejiang University, Zhejiang University, Hangzhou 310058, China*

**In memoriam Professor Jin Au Kong (1942-2008)**

**Featuring dense spatial distributions of engineered metallic particles, electromagnetic metamaterials exhibit simultaneously negative values of both, dielectric permittivity and magnetic permeability, within a resonance frequency band called left-handed passband. Unusual electromagnetic properties are found resulting in promising applications such as sub-wavelength resolution imaging. State-of-the-art micro/nanomanufacturing has led to resonance frequencies reaching the visible red. The common embedding of the metal particles in plastic matrices or deposition on dielectric substrates within a small area severely limits the usefulness of the materials. Here, we use UV or X-ray lithography to build**

comparably large areas and quantities of the first freely-suspended matrix-free metamaterials in which the metallic structures are S-string-like with their ends held by a window-frame. *In vacuo* spectral characterization combined with simulation reveals left-handed passbands from 1.6 to 2.2 THz. Owing to their size, the devices can be easily handled. They offer a straightforward way of making them tunable and two-dimensionally isotropic.

**Background**

After seminal work by Veselago[1] and Pendry and co-workers[2, 3], the field of electromagnetic metamaterials (EM$^3$) has soared with experimental demonstrations delivered in the GHz range[4- 13] and a wave of micro- and nanomanufacturing of THz devices including discrete elements[14- 21], hole arrays in thin metal films on both sides of a dielectric spacer[22], thin layers of dielectrics[23], vertical loops on a substrate[24], waveguide arrangements[25], and even a natural material based on colossal magnetoresistance[26]. Many interesting applications were proposed including sub-wavelength resolution imaging[27], invisibility cloaking[28, 29], and antennae[30, 31].

Usually, the metallic elementary units of electromagnetic metamaterials are either supported by a substrate or embedded in a matrix, both made from dielectric materials. Isolated small metallic elements such as split-ring resonators, U shapes or horseshoes, and rectangles[14- 21] have been implemented in the THz range while string-like elements such as S, $\Omega$, and dovetail were predominantly used in the microwave spectral range[9- 11]. All of them were held by plastic matrices or deposited on dielectric substrates. However, matrices and substrates may entail drawbacks for potential applications. Interaction of the incident wave and of the induced oscillations with the matrix may induce damping and shift of resonance frequency. EM$^3$ on a substrate may work only in reflection if the substrate is not transparent over the left-handed passband. Besides spectral absorption features, matrix materials such as photo or X-ray resists are polymeric and may suffer from degradation with age and radiation exposure. They are limited in operation temperature and potentially sensitive to water. Rigid substrates constrain potential applications. Finally, they do not offer a straightforward way of achieving tunability.

For the first time, we produce freely-suspended matrix-free large-area EM[3] by means of micromanufacturing. The basic idea is using string-like resonators and fixing the strings at both ends in a window frame. Here, we chose S resonators as investigated, in the microwave range, by J.A. Kong and co-workers[9-11]. Two window-frames are aligned and assembled in order to yield a bi-layer of strings that are aligned and separated by a gap of width d. While this window-frame is presently of polymeric material such as SU-8, for experimental convenience, numerical simulations show that it could as well be metallic. Besides making free-space devices, another important aspect of our work is to produce large quantities of good quality devices by means of UV and/or X-ray lithography, thereby restricting to the initial mask making the time-consuming primary pattern generation by e beam or laser beam writing. Finally, our micromanufacturing approach offers control of the resonance frequency over a wide range and leads straightforwardly to tunable and two-dimensionally isotropic devices.

The spectral response of our materials was characterized experimentally by Fourier Transform interferometry (FTIR) in the far infrared around 2 THz using synchrotron infrared radiation from the edge-effect source of the Helios 2 storage ring at SSLS. Measured spectra are compared with numerical simulations using commercial software MWS[32].

**Layout and specification**

For practical micromanufacturing, the basic layout of an S-string (fig. 1) is defined by the length a and width b of the S, the top and bottom width of the conductor h and h', its thickness t, and the interlayer gap d. Measured values of these parameters for the devices studied here are a=104.2 μm, b=34.9 μm, h=15.9 μm, h'=13.4 μm, and t=11.4 μm. The gap d has the values 0.6 μm, 1.1 μm, and 6.1 μm. Fig. 1 also shows the incoming wave and the incidence angle $\alpha$ as referred to the normal on the mid-plane of the bi-layer structure.

A window-frame made of SU-8 resist is holding S-strings by their ends, thus realizing a very precise regular array (fig. 2). The window-frame as built has four additional thin stabilizing ribs. The total size of chip and window is 14.7x12.3 mm$^2$ and 8.1x6.9 mm$^2$, respectively, the size of a sub-window 3.2x1.8 mm$^2$, and the thickness of the window-frame 260 μm.

**Micromanufacturing**

Although using two pieces of window-frame single-layer chips for one bi-layer chip, only one type of single layer needs to be manufactured for symmetry reasons. As a single-layer chip consists of a string layer 11.4 μm thick and the window-frame that is 260 μm thick, two lithography steps are performed into different resist depths. The bi-layer is then formed by assembling two complete single layers with their bottom sides facing each other. In this way, the alternating S structure that leads to resonance loops is automatically generated. Obviously, the assembly involves a spacer to define the gap d, a careful relative alignment of top and bottom layer, and a final glueing step to fix all assembled parts.

For the initial photomask, a 5" soda lime blank (Nanofilm, Wetlake Village, California) with 100 nm-thick chrome and 0.5 µm-thick layer of AZ1518 photoresist is used. A Heidelberg Instruments direct-write laser system DWL 66 serves for patterning the resist. Upon resist  development, chromium is etched away where it is not protected by the resist. Finally, the resist is dissolved leaving the thin chromium pattern that serves to absorb UV light.

Two distinct photomasks are needed, one containing the S-string array pattern, the other the window-frame pattern. 4×3 fields with either string or window-frame pattern are inscribed in a circle of 80 mm diameter such as to fit onto a standard 100 mm diameter Si wafer that serves as a substrate for the subsequent processing. Both masks have alignment marks for accurate relative positioning for the two exposures.

These photomasks are used either in a UV lithography process to expose structures in photoresist or to generate X-ray masks that feature thick ($\approx$20 µm) gold structures capable to absorb hard X-rays in case of X-ray deep lithography.

S-string arrays can be produced from both types of masks, photo as well as X-ray. The difference is that UV lithography using photomasks leads to sidewalls slanted by about 6° with AZ9260 resist while X-ray lithography delivers nearly ideal vertical sidewalls.

In the UV lithography process, S-string arrays are preferably patterned into a $\approx$20 µm thick AZ9260 resist layer which is spin coated on the silicon wafer that was covered with a plating base of Cr/Au (100/50 nm) before. After a soft baking at 95 °C for 3 min, the resist is exposed to UV light through the photomask by using a Karl Suss Mask & Bond Aligner (MA8/BA6). Finally, the exposed resist is removed with AZ 400k developer solvent and the remaining patterned resist structures serve as a mold for gold plating of the S-string arrays.

To make X-ray masks cost-effectively, a 100 mm diameter graphite wafer of 200 μm thickness is used as a membrane onto which a 35 μm thick SU-8 resist layer is spin coated. Photomask pattern are transferred to the SU-8 resist layer by UV exposure in the MA8/BA6 mask aligner. Post-exposure bake and development are followed by electroplating of 20 μm thick gold into the SU-8 mould. The graphite membrane with the patterned gold absorber is finally glued on an aluminium NIST-standard mask holder.

For X-ray lithography, the usual Si wafer with its plating base is covered with a ≈20 μm thick PMMA resist layer by multiple spin coating and soft baking. Each spin process yielding approximately 4 μm, five layers are required. The baking temperature is 180 °C, applied for about 2 min after each spin and 5 min after the final spin. The substrate is then exposed to synchrotron X-rays through an X-ray mask at the LiMiNT beamline of SSLS. Finally, exposed PMMA is removed by using the standard GG developer[33]. Thus formed PMMA structures serve for gold plating of the S-string arrays. Due to the short wavelength of X-rays used (≈0.2 nm) and the small local divergence of synchrotron radiation (≤1 mrad), sidewalls are smooth and nearly vertical.

In both cases, AZ or PMMA mold, a standard Enthone electrolyte is used for Au plating at a temperature of 50 °C. After electroplating, AZ resist is removed using a proprietary AZ remover whereas PMMA is dissolved by acetone.

For holding and stabilizing the freely-suspended matrix-free S-string array and for easily handling the final device, a window-frame is built that must be precisely aligned with the finished gold S-string array. Again, UV or X-ray lithography is used to expose this window-frame. First, SU-8 resist is spin coated to a thickness of 260 μm, followed by soft baking at 95 °C for 4 hours. Then, the sample is exposed either to UV light on the MA8/BA6 or to synchrotron X-rays at LiMiNT. Removing the unexposed SU-8 resist by SU-8 developer, we finally obtain ≈15 μm thick gold S-string arrays held by 260 μm thick SU-8 window-frames on a Si wafer.

To release these single layer chips from the Si substrate, the Cr layer of the plating base is used as a sacrificial layer to be etched away. For easy etching, holes are designed into the window-frame and the stabilizing ribs (fig. 3, (bottom)). First, the gold of the plating base is removed by submerging it for about 30 s into a KI solution (4g KI (potassium iodide) + 1 g $I_2$ (iodine) + 40 ml DI water). The then exposed Cr layer (100nm) is removed by chrome etchant CEP-200 (Microchrome Technology) effectively releasing the single layer chip from the substrate.

The assembly of two single-layer chips to a bi-layer chip is the remaining step. Relative alignment of the S-string arrays in both single-layer chips and the control of the gap between two layers are critical issues.

The gap is set by sandwiching a spacer made from SU-8 along the processes described above. The shape of the spacer corresponds to that of the window-frame. Different thicknesses ranging from 4 μm to 20 μm are fabricated to control mostly capacitance and ensuing pass-band frequency. The alignment is done by means of a special optical microscope equipped with micrometric translation and rotation features. When the alignment is satisfactory, the sandwich of two single-layer chips with the spacer foil in between is fixed by glueing to obtain the final product. Figs. 3 and 4 show the bi-layer chip with freely-suspended matrix-free S-string arrays in various magnifications. Images were taken by means of a digital camera (fig. 3), an optical microscope, and a scanning electron microscope (fig. 4).

**Spectral characterization**

For spectral characterization, the bi-layer chip is positioned on a rotational stage in the sample chamber of a Bruker IFS 66 v/S Fourier transform interferometer. The rotation axis is parallel to the string direction as is the electric field vector of the incident radiation. Synchrotron radiation at the ISMI beamline of SSLS is used as a source. The beam spot on sample is about 1 mm wide horizontally and 0.4 mm vertically. A far infrared (FIR) polarizer (gold grid on polyethylene substrate) is introduced before the sample. Transmission spectra are acquired by means of a DTGS detector comparing signals with and without samples. The spectra are taken from nearly 1 THz up to 10 THz with a resolution of 0.12 THz (4 cm$^{-1}$).

The incidence angle α is varied from normal incidence 0° up to 60°. In this way, the magnitude of the magnetic field component perpendicular to the inductance loop is varied. Using the transmission of SU-8 in the THz range as measured earlier[17], namely, 0.93 for 25 μm at 1 THz, we estimate that one leg of the window-frame would have a transmission of $3.9 \cdot 10^{-4}$, thus ruling out measurements under 90° incidence angle.

The sample chamber is evacuated to a level of 4 mbar of dry nitrogen gas in order to avoid signal contributions from water vapour and other gases.

The geometrical dimensions of the bi-layer chips as built are measured by means of a scanning secondary electron microscope (SEM, FEI Sirion), an optical profiler (WYKO) and an optical microscope (Leica).

**Measured and simulated spectral results**

Fig. 5 compares three pairs of measured and simulated transmission spectra which differ only by the gap width. The geometrical parameters are the same for all three samples as given above, except for the interlayer gap which is 0.6 μm, 1.1 μm, or 6.1 μm wide. For simulation, CST's MWS package is used[32].

The basic features of the measured spectra are reproduced by the simulated ones. In particular, in the case of normal incidence (α=0°), there are two prominent pass bands measured at 2.2 THz and between 2.6 to 2.8 THz which can be found as peaks at 2.1-2.4 THz and 2.7-3.6 THz in the simulation. The apparent difference in frequency may be caused by errors in the measured parameters of the bi-layer chips, in particular, the gap. The dependence of the peak position on incidence angle is also well reproduced for the two smaller values of the gap (0.6 μm, 1.1 μm), whereas it deviates more significantly for the large gap of 6.1 μm. This angular dependency is related to the fat aspect of the conductors leading to spatial dispersion because location and size of a resonance loop as defined by the actual current flow may vary with the incidence angle, thus changing inductance and capacitance.

Finally, the big peak at 90° incidence angle shown by the simulation cannot be observed experimentally due to the complete obstruction by the window-frame. However, there is a weak indication that this peak starts developing at smaller incidence angles already. We note that the window-frame can be built, for future experiments, such as to allow experimental access to 90° incidence.

From a robust retrieval algorithm[34], values of the effective permittivity ε and permeability μ are obtained which show that the lower frequency pass band around 2.2

THz is left-handed involving a magnetic resonance (fig. 6). This finding was further confirmed by phase tracking that showed backward wave propagation inside the bi-layer. In contrast, the higher frequency pass band between 2.7-3.6 THz is an electric resonance with both ε and μ positive and thus right-handed.

In this normal incidence case, the magnetic field vector is parallel to the x axis. Inductive coupling to the S-strings is possible nonetheless because the inductance loop that is formed by legs b, $\frac{a}{2}$, and b in the top and, with opposite orientation, in the bottom layer lies in a plane inclined by an angle of $\tan \varphi = (d+t)/(b-(h+h^t)/2)$ which is about 30° with respect to the xz plane for the gap d=0.6 μm. From one half-S loop to the next, the sign of this inclination angle alternates. Furthermore, as the spatial extension of the conductors is not negligible to their distances, there are many possible induction loops distributed over a certain angular range around $\varphi$. If $\alpha$ is the incidence angle of the wave on the xz mid-plane, the magnetic field components perpendicular to each loop are $H_0 \sin(\pm \varphi - \alpha)$ with the sign alternating from loop to loop. This alternating inclination of loop surfaces is actually an advantage of the bi-layer S structure as it provides a non-vanishing magnetic coupling at any incidence angle $\alpha$.

We note that, for $|\alpha| \leqslant |\varphi|$, and in particular $\alpha = 0$, the current flows in one direction along the whole string whereas for $|\alpha| > |\varphi|$ the current flow changes direction from loop to loop which we may call a string resonance and a loop resonance, respectively. A visible sign of these different resonance modes may be the strong reduction of signal amplitude at 30° and higher as compared to 15° and 0°.

In addition to this resonance loop between a top and bottom string at the same x position, there is another resonance loop formed by leg a/2 of a bottom string and leg a/2 of the adjacent top string. Its centerline area is $A = \frac{a}{2} \cdot \sqrt{(d+t)^2 + (p_w - (b - \frac{h+h'}{2}))^2}$ which is smaller than the main loop area of $ab/2$ for our geometric parameters and the inclination angle $\tan \chi = (d+t)/(p_x - (b - \frac{h+h'}{2}))$, about 35° for the same structure as above. A similar consideration as for the preceding loop would apply for this interstring loop. It would be expected to have a smaller inductance as well as a smaller capacitance, so a higher frequency.

Furthermore, we note that the design of the bi-layer chips opens an immediate way of achieving tunability. Replacing the fixed thickness spacer by devices whose thickness can be controlled such as piezoelectric actuators, spring-loaded capacitive microactuators or shape-memory alloys would allow controlling the gap width d and the resonance frequency, especially in the 90° incidence case.

Finally, the left-handed behavior in x and y directions as shown in fig. 6 implies an easy realization of a two-dimensionally isotropic metamaterial by optimising S-string parameters, in particular, the gap d. This feature seems inaccessible to other THz devices as published so far.

**Conclusion**

The micro/manufacturing of electromagnetic metamaterials based on precisely aligned

freely-suspended matrix-free bi-layers of S-strings over comparably large areas up to 56

mm² is demonstrated for the first time. The devices actually built exhibit left-handed pass bands around 2.2 THz. This development opens up applications for large-area THz electromagnetic metamaterials in fields including infrared optics and imaging. Still larger device areas will be accessible via possible extensions of the process. As micro/nanomanufacturing allows a variation of the geometry of the strings within a broad range, resonance frequencies can be adapted to values as required by applications. Furthermore, two-dimensionally isotropic materials may be built and, finally, devices could be made tunable by including means to vary the gap width.

## Acknowledgements

Work partly performed at SSLS under DARPA HR0011-06-1-0030, NUS Core Support C-380-003-003-001, A*STAR/MOE RP 3979908M and A*STAR 12 105 0038 grants. JAK and HSC would also like to acknowledge the Chinese National Science Foundation under contract 60531020.

Correspondence and Requests for materials should be addressed to: Moser H. O. (moser@nus.edu.sg)

**Figure Legends**

Figure 1  Basic layout of the S string and the geometry of the incident wave. S string is drawn to scale with respect to the actually fabricated strings. Strings repeat periodically in x direction with a periodicity constant $p_x$

Figure 2  Schematic design of bi-layer chip. (a) Drawing of a sheet of aligned bi-layer strings 5a long and $10p_x$ wide. (b) Schematic of bi-layer chip showing top and bottom window-frame as well as the spacer foil.

Figure 3  Micromanufactured bi-layer chip. (a and b) Tweezer and ruler illustrate the handling and size of the device. (c) Bird's eye view of about $1.2 \times 0.9$ mm$^2$. A slight defect can be seen in the fifth column from the right where the top string is bulging a little to the right. Scale bar is 250µm.

Figure 4  Various views of a bi-layer chip taken with optical microscope and scanning electron microscope SEM. (a) Optical microscope focusing on the bottom layer. (b) Optical microscope focusing on the top layer. For both (a) and (b) the scale bar is 50 µm. (c) SEM bird's  eye view (scale bar 200 µm. (d) SEM close-up (scale bar 50µm). All pictures show the accurate definition of the S-string bi-layers.

Figure 5  Measured (a, b, c) and simulated (d, e, f) transmission spectra of three bi-layer chips having parameters a=104.2 µm, b=34.9 µm, h=15.9 µm, h'=13.4

µm, and t=11.4 µm. The gap d has the values 0.6 µm (a, d), 1.1 µm (b, e), and 6.1 µm (c, f). The parameter denoting the individual curves is the incidence angle $\alpha$ as measured with respect to the normal on the bi-layer midplane.

Figure 6   Calculated effective permittivity $\varepsilon_z$ and permeabilities $\mu_x$ and $\mu_y$ for the sample with gap d=6.1 µm for 0° (a) and 90° (b) incidence.

1   **Figures**

2   [Insert Figures here.]

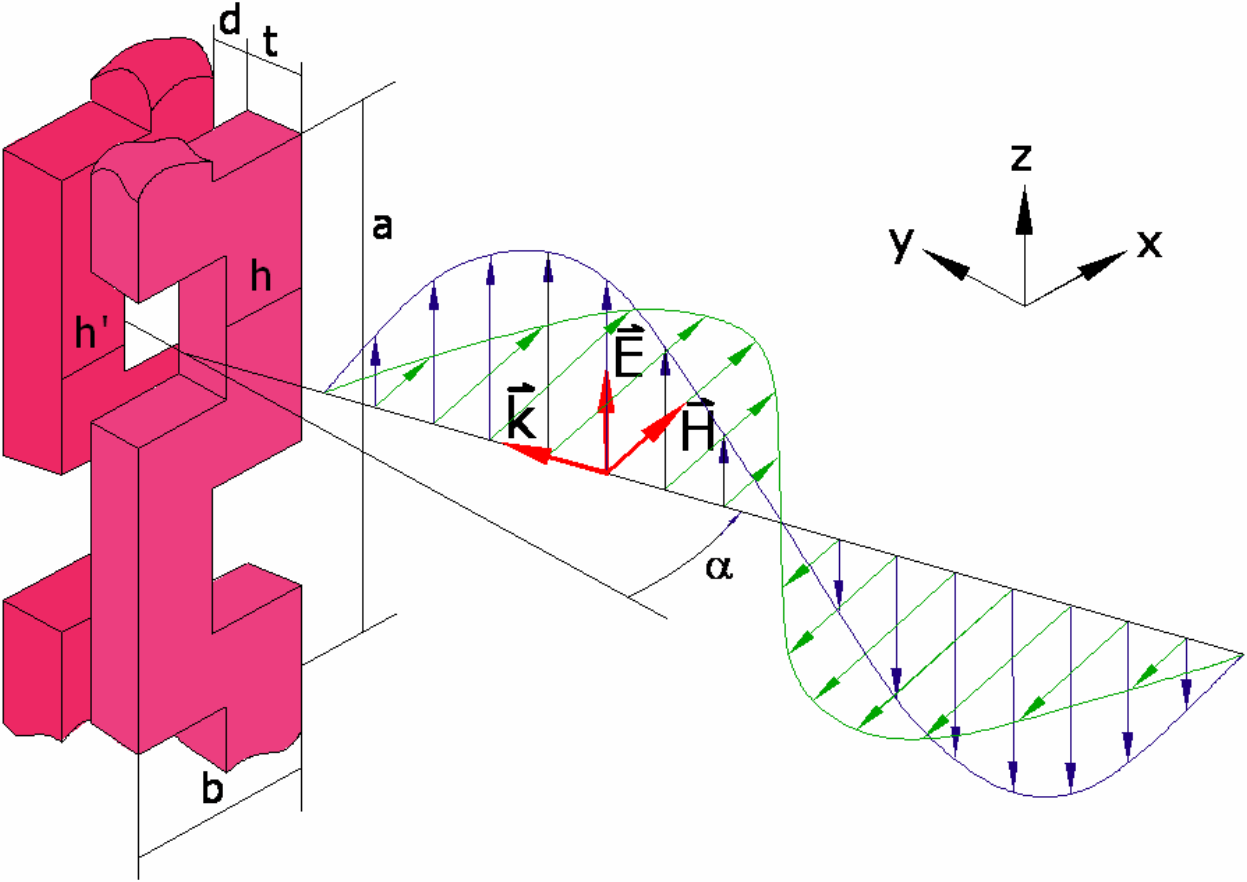

Freely-suspended matrix-free metamaterials

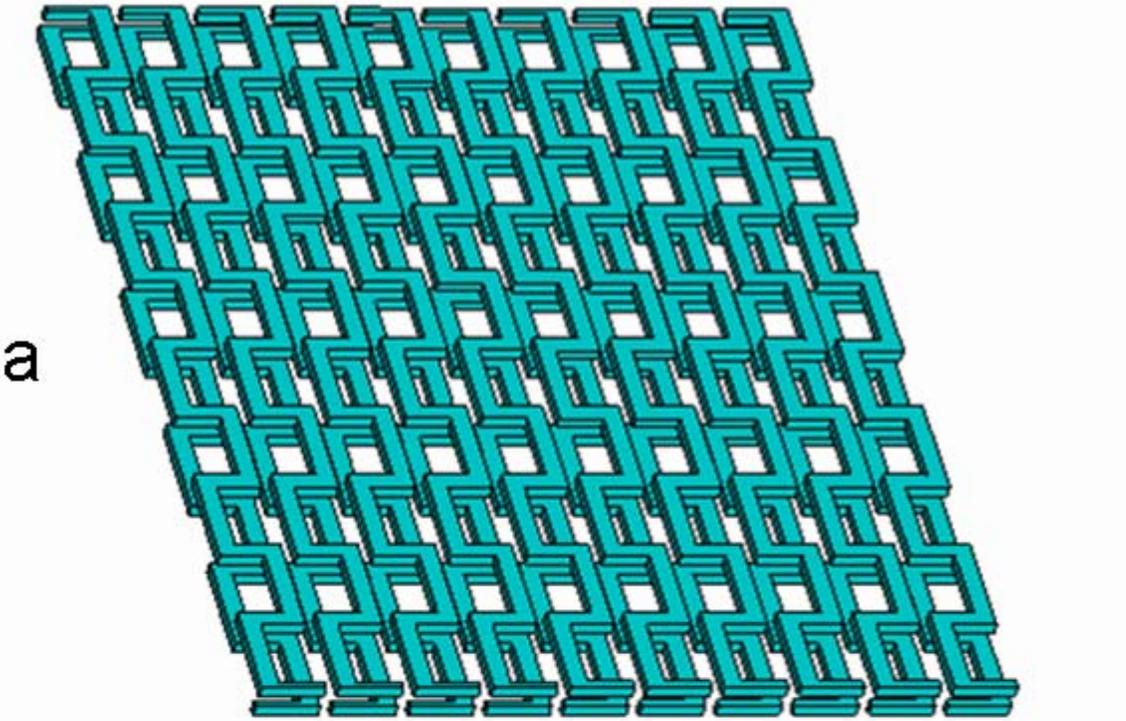

a

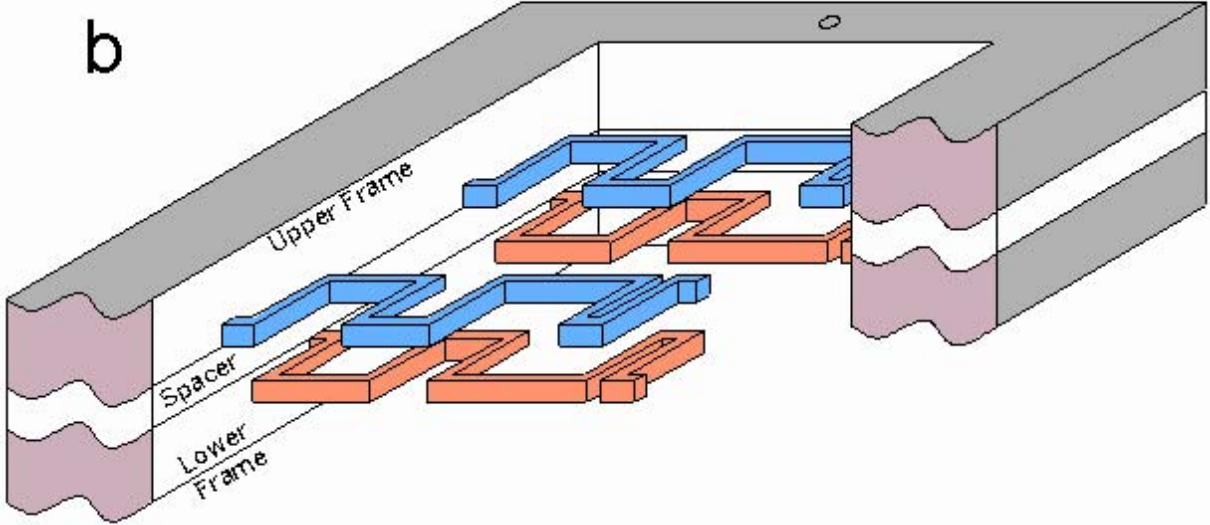

b

Freely-suspended matrix-free metamaterials

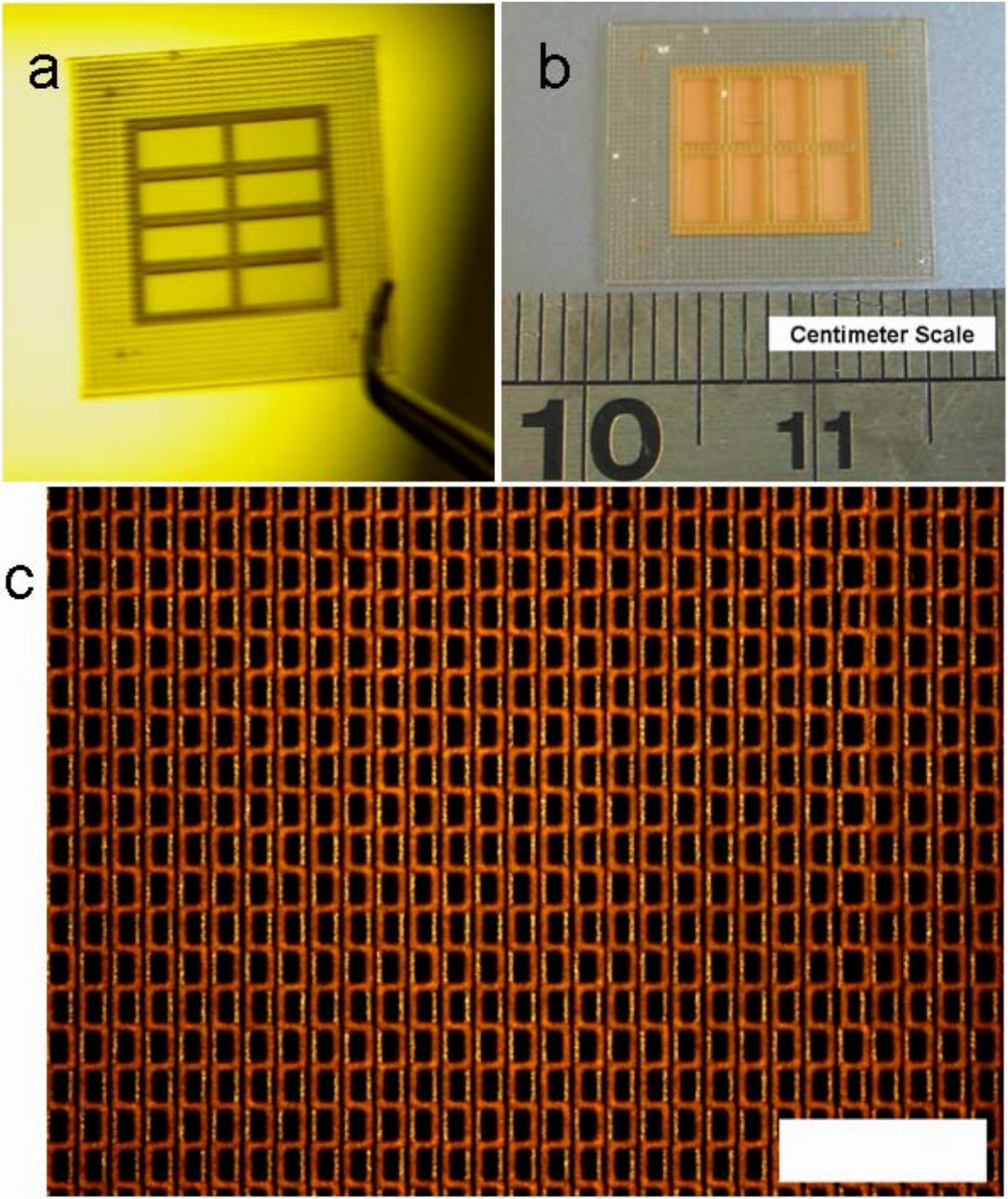

Freely-suspended matrix-free metamaterials

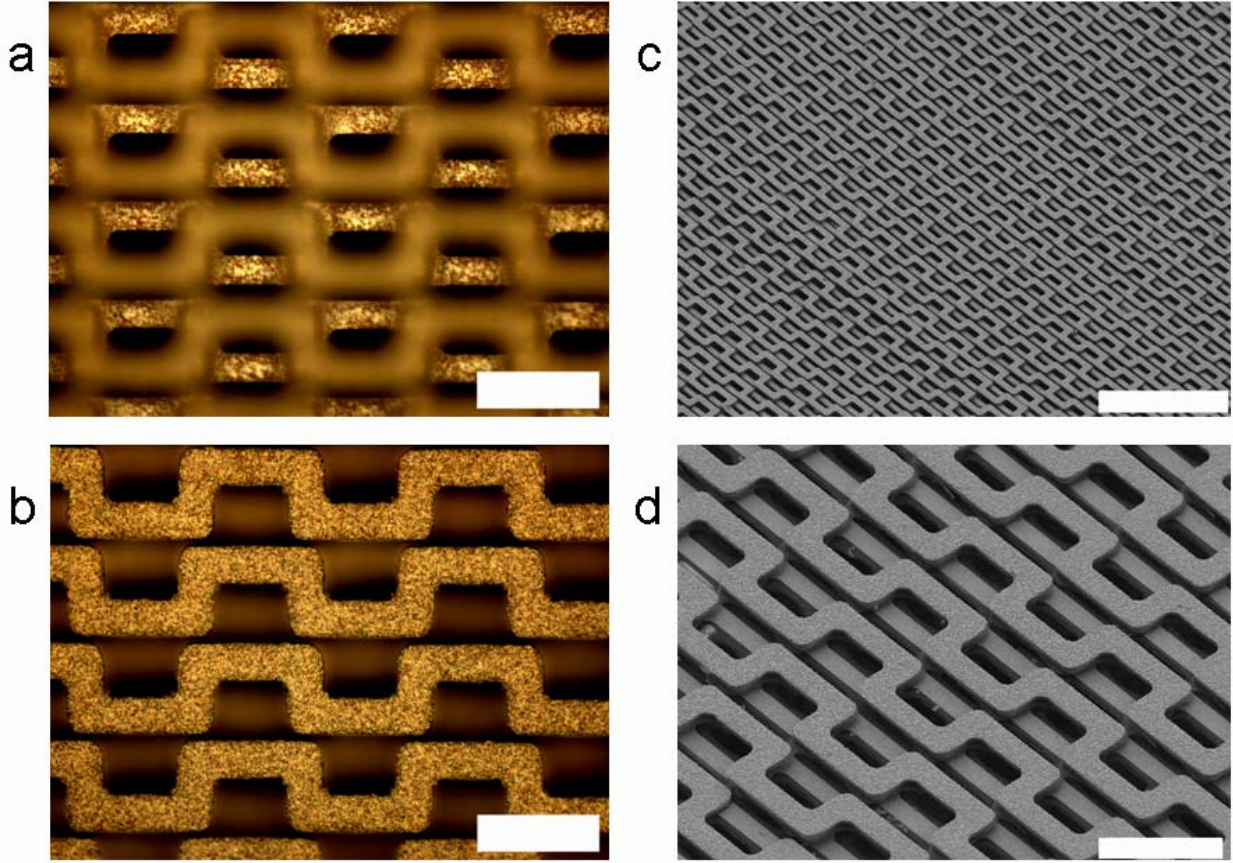

Freely-suspended matrix-free metamaterials

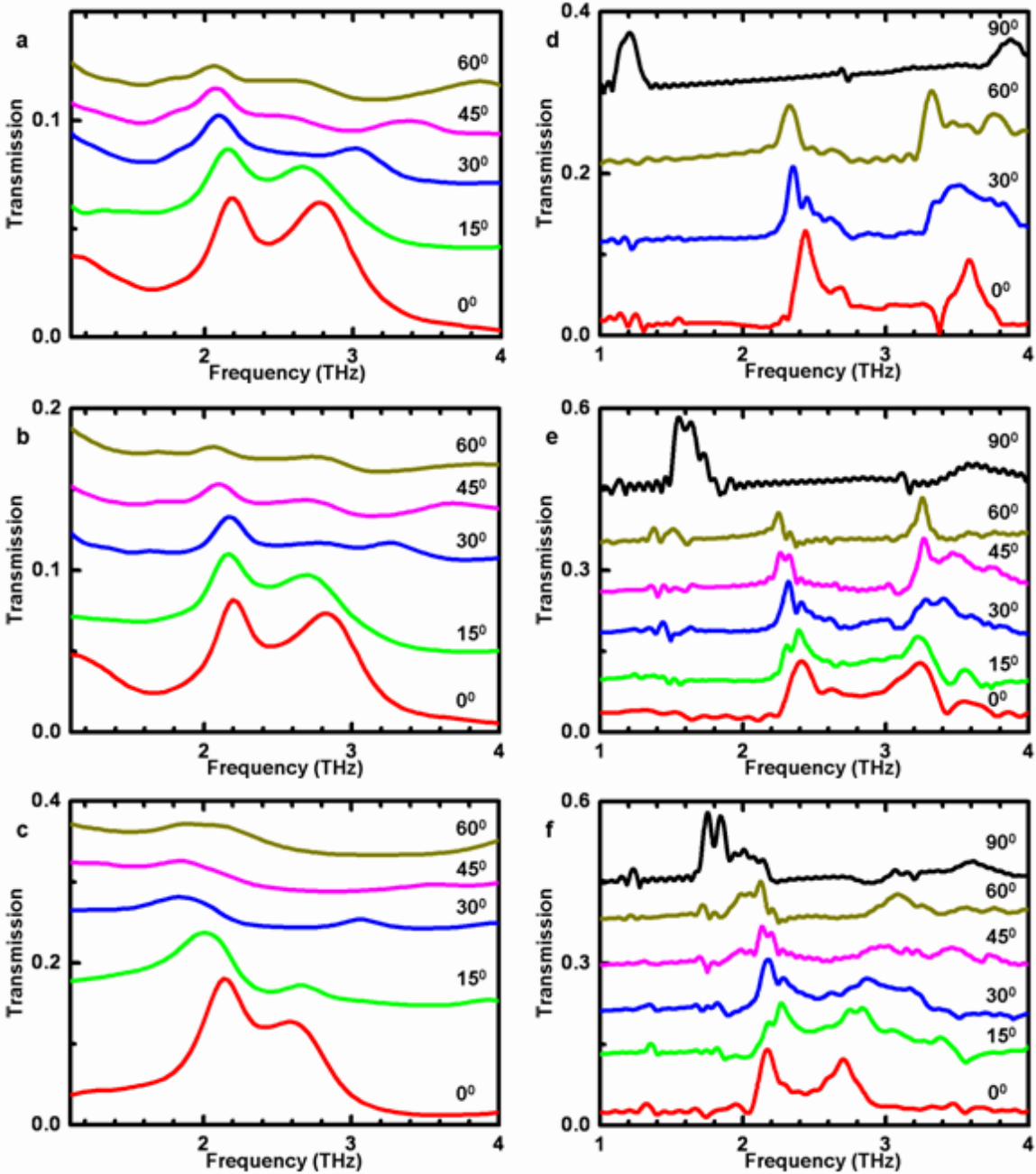

Freely-suspended matrix-free metamaterials

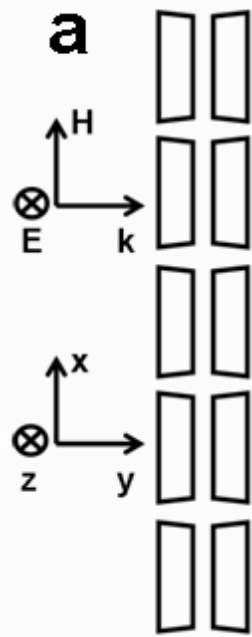
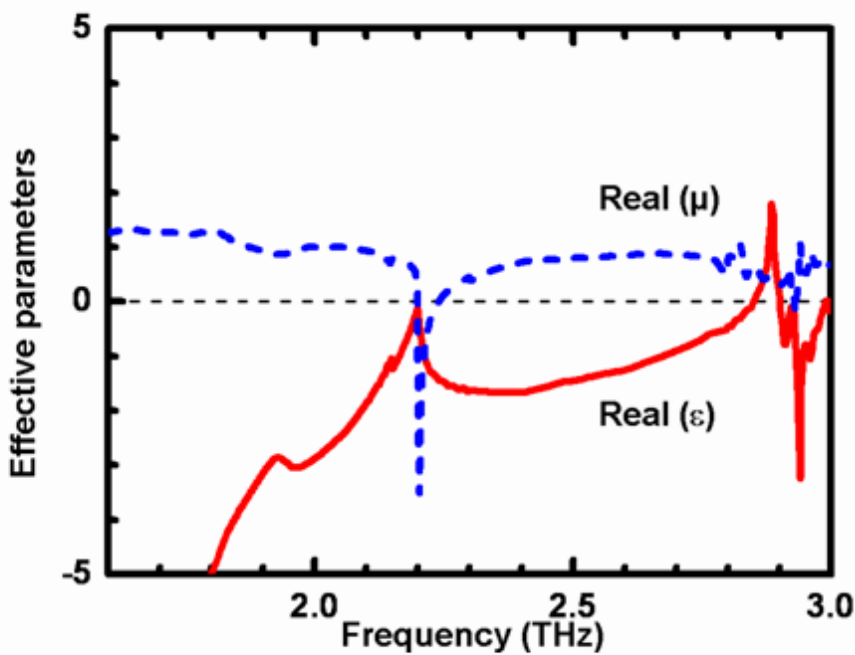

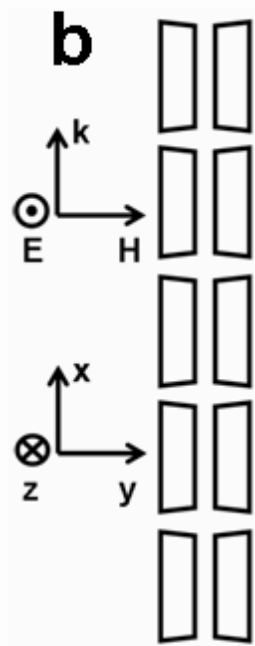
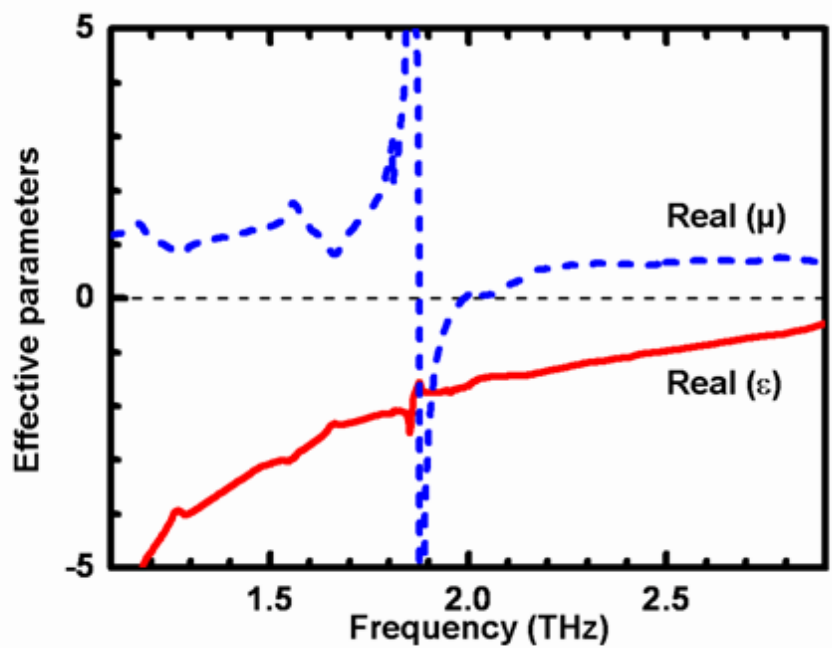

Freely-suspended matrix-free metamaterials